\documentstyle[12pt]{article}
\advance\textheight by \topskip
\begin{document}
\thispagestyle{empty}
\begin{flushright}
CERN-TH/95-15\\
DFTT 5/95\\
SUSSEX-AST-95/2-1\\
IEM-FT-101/95\\
astro-ph/9502010\\
March 1995
\end{flushright}
\vskip 8mm
\begin{center}
{\Large\bf Astrophysical and cosmological constraints\\
\vskip .5cm
on a scale-dependent gravitational coupling}
\vskip 8mm
{\bf Orfeu Bertolami}\footnote{On leave of absence from
Departamento de F\'\i sica, Instituto Superior T\'ecnico, Av. Rovisco Pais,
1096 Lisboa Codex, Portugal; also at Theory Division, CERN;
E-mail: Orfeu@vxcern.cern.ch}
\vskip .1mm
{\it INFN - Sezione Torino, Via Pietro Giuria 1, I-10125 Torino, Italy}
\vskip 5mm
{\bf Juan Garc\'\i a--Bellido}\footnote{PPARC postdoctoral research
fellow. E-mail: j.bellido@sussex.ac.uk}
\vskip .1mm
{\it Astronomy Centre, University of Sussex, Brighton BN1 9QH, UK}
\end{center}
\vskip 8mm

{\centerline{\Large\bf Abstract}
\begin{quotation}
\vskip -0.4cm
We study the phenomenological consequences of the recently proposed idea
of a running gravitational coupling on macroscopic scales. When applied
to the rotation curves of galaxies, we find that their flatness requires
the presence of baryonic dark matter. Bounds on the variation of the
gravitational coupling from primordial nucleosynthesis and the change of
the period of binary pulsars are analysed. We also study constraints on
the variations of $G$ with scale from gravitational lensing and the
cosmic virial theorem, as well as briefly discuss the implications of
such a scenario for structure formation.
\end{quotation}

\newpage

\section{\label{1}Introduction}

The flatness of the rotation curves of galaxies and the large structure
of the Universe indicate that either the Universe is predominantly made
up of dark matter of exotic nature, i.e. non-baryonic, and/or that on
large scales gravity is distinctively different from that on solar
system scales, where Newtonian and post-Newtonian approximations are
valid. The former possibility has been thoroughly investigated on
astrophysical as well as on cosmological grounds (see Ref.~\cite{Trimble}
for a review) and is one of the most active subjects of research in
astroparticle physics. The second possibility, however relevant, has drawn
little attention so far. This
may be essentially due to the fact that until recently no consistent and
appealing modification of Newtonian and post-Newtonian dynamics has been put
foward. Many of these attempts \cite{Finzi},
although consistent with observations, were most
often unsatisfactory from the theoretical point of view. Actually, it has
been recently shown that under certain fairly general
conditions it seems unlikely that
relativistic gravity theories can explain the flatness of the rotation curves
of galaxies \cite{Nester}.
These conditions however do not exclude the class of
generalizations of General Relativity that involve higher-derivatives.
Quantum versions of these theories were shown to exhibit asymptotic
freedom in the gravitational coupling \cite{Julve} and one would expect
this property to manifest itself mainly on large scales. This possibility
would surprisingly imply that quantum effects could actually mimic the
presence of dark matter \cite{Goldman}, as well as induce other
cosmological phenomena \cite{Bertolami,Bertolami1}.} One of the
most striking implications of these ideas is the prediction that the
power spectrum on large scales would have more power than the one predicted
by the $\Omega = 1$ Cold Dark Matter (CDM) Model, in agreement with what is
observed by IRAS \cite{IRAS}. Furthermore, due to the increase in the
gravitational constant on large scales one finds that
the energy density fluctuations grow quicker than in usual
matter dominated Friedmann-Robertson-Walker models. Moreover, one
can naturally explain with a scale-dependent $G$ the discrepancy between
determinations of the Hubble's parameter made at different scales,
as suggested in \cite{Bertolami}, and recently studied in \cite{Kim}.

Nevertheless, independently of the possible running of the gravitational
constant in a higher derivative theory of gravity, it is certainly
worthwhile analysing the constraints on the scale-dependence of $G$ from
astrophysical and cosmological phenomena, where such an effect would
be dominant. On the other hand, in the last few years there has been a
revival of Brans-Dicke like theories, with variable gravitational coupling,
that has led to a number of phenomenological constraints on possible time
variations of $G$. Of course, some of the constraints on $\dot G$ can be
written as constraints on $\Delta G$ over scales in which a graviton
took a time $\Delta t$ to propagate. For instance, during
nucleosynthesis the largest distance that a graviton could have
traversed is the horizon distance at that time, {\it i.e.} a
few ligh-seconds to a few light-minutes, or $10^{10}$ to $10^{12}$ cm,
approximately the Earth-Moon distance. Such a distance is too small
for quantum effects to become appreciable, as we will discuss below.
However, those effects become important at kiloparsec (kpc) distances
and therefore could be relevant for discussing the rotation curves of
galaxies. We shall actually show, for a particular theory
\cite{Goldman,Bertolami}, that
the rotation curves of spiral galaxies cannot be entirely explained by
the running of $G$, so some amount of baryonic dark matter is required,
which is still consistent with the upper bound on baryonic matter coming
from primordial nucleosynthesis. This result is generic of theories
with a power-law dependence of the gravitational coupling on scale.
On the other hand, we could impose bounds on a possible variation of $G$
from a plethora of cosmological and astrophysical phenomena at large
scales, although the lack of precise observations at those scales
will make the bounds rather weak. It is nevertheless
expected that the increasing
precision of future experiments might tell us something about variations
in $G$. However, since for most of these phenomena the gravitational
constant appears in the factor $GM$, we cannot actually distinguish a
variation in $G$ from the existence of some peculiar kind of dark matter.
In fact, dark matter seems to us like some kind of `ether', an {\it ad hoc}
and unobserved medium that permeates space modifying the behaviour of
otherwise well known baryonic matter. If the idea of an asymptotically
free gravitational coupling is correct, we might be able to get rid of
this elusive yet dominating component of the Universe.

Furthermore, a scale dependence of the gravitational constant arises
from completely different reasons in the stochastic inflation formalism,
as recently explored in \cite{LLM} and \cite{GBL}. The scaling behaviour
and screening of the cosmological constant was also discussed in the
context of the quantum theory of the conformal factor in four dimensions
\cite{AMM}, as the theory approaches its infrared fixed point, at
distance scales much larger than the horizon size.
The way the gravitational constant varies with scale in each case is
very different from that of the asymptotically free theories, so it
seems worth studying the phenomenological constraints that might rule
out one or another.

\section{\label{2}Asymptotic Freedom of the Gravitational Coupling}

The main idea behind the results of Refs. \cite{Goldman,Bertolami}
is the scale depedence of the gravitational coupling.
The inspiration for this comes from the property of asymptotic freedom
exhibited by 1-loop higher--derivative quantum gravity models \cite{Julve}.
Since there exists no screening mechanism
for gravity, asymptotic freedom may imply that quantum  gravitational
effects act on macroscopic and even on cosmological scales, a fact which
has of course some bearing on the dark matter problem
\cite{Goldman} and in particular on the large scale structure of the
Universe \cite{Bertolami,Bertolami1}. It is within this framework that
a power spectrum which is consistent with the observations of IRAS
\cite{IRAS} and COBE \cite{COBE} can be obtained and where energy
density fluctuactions are shown to grow faster than in usual
cosmological models \cite{Bertolami,Bertolami1}.
This last feature does bring some hope that
the large scale structure may arise from primordial energy density
fluctuations entirely amplified by an asymptotically
free gravitational coupling and baryonic matter.

Let us now briefly outline this proposal.
Removing the infinities generated by quantum fluctuations and ensuring
renormalizability of a quantum field theory requires a scale--dependent
redefinition of the physical parameters. This was done at the 1-loop
level in a higher--derivative theory of gravity in Refs. \cite{Julve}.
Furthermore, the removal of those infinities still leave the physical
parameters with some dependence on finite quantities whose particular
values are arbitrary. These can be assigned by specifying
the value of the physical parameters at some momentum or length scale;
once this is performed, variations on scale are accounted for by
appropriate changes in the values of the physical parameters as
described by the renormalization group equations (RGEs). Thus, the
equations of motion in the quantum field theory of gravity should be
similar to the ones of the classical theory, but with their parameters
replaced by the corresponding `improved' values, that are solutions
of the corresponding RGEs. However, since gravity couples coherently to
matter and exhibits no screening mechanism, quantum fluctuations of the
gravitational degrees of freedom contribute on all scales. One must
therefore include the effect of these quantum corrections into the
gravitational coupling, $G$, promoting it into a scale--dependent
quantity. One-loop quantum gravity models indicate that the coupling
$G (\mu^2 / \mu_*^2  \sim r_*^2/r^2)$ is asymptotically free, where
$\mu_*$ is a reference momentum, meaning that $G$ grows with scale
\cite{Julve}. A typical solution for $G(r_*^2 / r^2)$ was obtained in
Ref.~\cite{Goldman} setting the $\beta$-functions of matter to vanish
and integrating the remaining RGEs:
\begin{equation}\label{GR}
G(r_*^2 / r^2)=G_{lab}(r_*^2 / r_{lab}^2)\, \delta(r,r_{lab})\ ,
\end{equation}
where $G_{lab}(r_*^2 / r_{lab}^2)$ is the value of $G$ measured
in the laboratory at a length scale $r_{lab}$, and
$\delta(r,r_{lab})$ is a growing function of $r$,
such that it is equal to one at $r = r_{lab}$.
A convenient choice for $r_*$ is $r_* = r_{lab}$. In order for the
asymptotic freedom of $G(\mu^2 / \mu_*^2)$ to have an effect
on for instance the dynamics of galaxies and their rotation curves,
the function $\delta(r,r_{lab})$ should be close to one for $r <
1$ kpc, growing significantly only for $r \geq 1$ kpc.
Naturally, this dependence of $G$ with distance has also
implications of cosmological nature.
A convenient parametrization for $\delta(r,r_{lab})$ from
the fit of Ref.~\cite{Goldman} in the kpc range is the following:
\begin{equation}\label{DEL}
\delta(r,r_{lab}) = 1.485 \left[1 + \beta \left({r \over r_0}\right)^{\gamma}
\ln{r \over r_0}\right]\ ,
\end{equation}
where $\beta \simeq 1/30$, $\gamma \simeq 1/10$ and $r_0 = 10$ kpc.

We shall use this fitting in the next Section in our analysis
of the rotation curves of galaxies, and extract from it a prediction
for the distribution of baryonic dark matter. However, before we pursue
this discussion let us present some of the ideas developed in Refs. [5-7].
As discussed above, the classical equations
have to be `improved' by introducing the scale dependence of the
gravitational coupling. This method suggests that the presence of
cosmological dark matter could be replaced by an asymptotically free
gravitational coupling. Assuming that the Friedmann equation describing
the evolution of a flat Universe is the improved one, then:
\begin{equation}\label{H2}
H^2(\ell) = {8 \pi \over 3} G(\frac{a_0^2
 \ell_*^2}{a^2 \ell^2}) \rho_{m}\ ,
\end{equation}
where $a=a(t)$ is the scale factor, $H=\dot{a}/a$ is the Hubble
parameter, $\rho_m$ is the density of baryonic
matter, $\ell$ is the comoving distance and
$\ell_*$ is some convenient length scale.

{}From Eq.~(\ref{H2}) one sees that by construction the
present physical density parameter, $\Omega_0^{phys}$, is equal
to one. However, the quantity which is usually referred to as density
parameter is actually:
\begin{equation}\label{OME}
\Omega_0 = {8 \pi \over 3} {G \rho_{m_0} \over {H_{0_*}^2}}\ ,
\end{equation}
where $H_{0_*}$ is the present Hubble parameter for a
given large scale distance, $r = r_*$, and the value
of the product $G \rho_{m_0}$ is inferred for different scales
with the help of the Virial Theorem. This leads to
$\Omega_0$ as a growing function of scale, which is in
agreement with observations for a constant $\rho_{m_0}$. We hence
conclude that macroscopic quantum gravity effects do
mimic the presence of dark matter.

Furthermore, from Eq.~(\ref{H2}) one can clearly see the scale
dependence of the Hubble parameter \cite{Bertolami,Bertolami1,Kim}.
In view of the above arguments, the criticism raised in
Ref.~\cite{BKS}\footnote{We thank A. Bottino to call our attention to
this reference while we were writing our paper.} concerning the way
the effect of the running of the gravitational coupling on $\Omega_0$
was considered in Ref.~\cite{Bertolami}, seems to be unjustified
since, as explained above, $\Omega_0^{phys} = 1$. Moreover, as shown
in Refs. \cite{Bertolami} and \cite{Bertolami1}, the power spectrum
resulting from these considerations is similar to that of a low
density Cold Dark Matter model with a large cosmological constant
\cite{Efs1} which in terms of the density matter:
\begin{equation}\label{CDM}
\Omega_0^{CDM}=0.15 \qquad \Omega_0^{\Lambda}=0.80 \qquad
\Omega_0^{Baryons}=0.05\ .
\end{equation}

Another popular, although rather {\it ad hoc},
possibility to account for the
large scale structure of the Universe consists of a mixture of
Cold and Hot Dark Matter \cite{Davis}:
\begin{equation}\label{HDM}
\Omega_0^{CDM}=0.65 \qquad \Omega_0^{HDM}=0.30 \qquad
 \Omega_0^{Baryons}=0.05\ .
\end{equation}

Although these two last possibilities are compatible with COBE
\cite{COBE} and IRAS \cite{IRAS} data one could consider an
alternative model where the gravitational coupling is
scale-dependent. We stress that the recently discovered evidence for
gravitational microlensing by massive astrophysical objects in the
Galactic halo \cite{Alc,Aub} implies, at least preliminarly, that a
sizeable fraction of the halo is composed by non-luminous {\it
baryonic} matter. This represents a serious difficulty for the
existing models of structure formation since baryonic dark matter is
notoriously inefficient as to the amplification of energy density
fluctuations. Furthermore, the estimated ratio in density of
non-luminous baryonic matter to cold dark matter cannot exceed in
those models $1/4$ and $1/10$, respectively, and reported
gravitacional microlensing events by the EROS collaboration \cite{Aub}
are compatible with the halo being entirely composed by Massive
Compact Halo Objects (MACHOs). Thus, if it turns out that MACHOs do
indeed dominate the halo, then it becomes particularly important to
look for alternatives to the existing structure formation models.

\section{\label{3} Rotation Curves of Galaxies}

Let us now turn to the discussion of the implications of the fit
(\ref{DEL}) for the rotation curves of galaxies. It is a quite well
established observational fact that the rotation curves of spiral
galaxies flatten after about 10 to 20 kpc from their centre, which of
course is a strong dynamical evidence for the presence of dark matter
and/or of non-Newtonian physics. The rotation velocity of the galaxy
is given by the non-relativistic relation,
\begin{equation}\label{V2}
v^2(r) = {G(r)M(r)\over r}\ ,
\end{equation}
which approaches a constant value some distance from the centre,
{\it e.g.} $v_0^2 = 220$ km/s for our galaxy. Assuming that the
gravitational coupling is precisely Newton's constant $G_N$ and
imposing that the rotation velocity is constant, using the Virial
Theorem at $r = R \equiv 500$ kpc, one finds the standard expression
for the mass distribution of dark matter:
\begin{equation}\label{MNR}
M_N(r) = M_N(R) {r \over R}\ .
\end{equation}
Assuming instead a running gravitational coupling (\ref{DEL}), the
condition that the rotation velocity is constant yields:
\begin{equation}\label{MR}
M(r) = {0.673 \over \left[1 + \beta ({r\over r_0})^{\gamma}
\,\ln{r\over r_0}\right]} M_N(r)\ .
\end{equation}
Equation (\ref{MR}) reveals after simple computation that the running
of the gravitational coupling reduces the amount of dark matter
required to explain the flatness of the rotation curves of galaxies by
about $44 \%$, assuming that galaxies stretch up to a distance of
about 500 kpc. This result is in agreement with Ref.~\cite{BKS}, a
clear prediction of the dependence of the gravitational coupling with
scale and, in particular, of the fit (\ref{DEL}).  Furthermore, since
the possibility that the Galactic halo is entirely made up of baryonic
dark matter is barely consistent with the nucleosynthesis bounds on
the amount of baryons \cite{CST}, the running of $G$ is quite welcome
since it reduces the required amount of baryonic dark matter in the
halo (although not in the bulge). An entertaining hypothesis could be
that precisely this effect is responsible for the reduction in the
microlensing event rates across the halo in the direction of the Large
Magellanic Cloud with respect to those along the bulge of our galaxy,
as reported by
\cite{Udalski}.

\section{\label{4} Bounds on the variation of $G$ with scale}

In this section we constrain the variation of the gravitational coupling
given by the fit (\ref{DEL}) with bounds from primordial nucleosynthesis,
binary
pulsars and gravitational lensing. We shall also discuss the effect that
a scale-dependent $G$ has on the peculiar velocity field and how future
experiments might help resolving such an effect at cosmological distances.

\subsection{Primordial nucleosynthesis}

As mentioned in the introduction, one could obtain bounds on the
variation of the gravitational coupling from observations of the light
elements' abundances in the Universe. Such observations are in
agreement with the standard primordial nucleosynthesis scenario
(for a review see \cite{OSSW}), but there is still some room for
variations in the effective number of neutrinos, the baryon fraction
of the universe and also in the value of the gravitational constant.
For instance, the predicted mass fraction of primordial $^4${\it He}
can be parametrised, in theories with a variable gravitational coupling,
in the following way \cite{OSSW,CGQ},
\begin{equation}\label{PNS}
Y_{\rm p} = 0.228 + 0.010 \ln \eta_{10} + 0.327 \log \xi\ ,
\end{equation}
where $\eta_{10}$ is the baryon to photon ratio in units of $10^{-10}$
and $\xi$ is the ratio of the Hubble parameter at nucleosynthesis and
its present value, itself proportional to the square root of the
corresponding gravitational constant. In the fit
(\ref{PNS}) we have assumed that the effective number of light neutrinos
is $N_\nu = 3$  and that the neutron lifetime is $\tau_n =
887$ seconds \cite{PDG}.

By running the nucleosynthesis codes for different values of $G$,
Accetta, Krauss and Romanelli \cite{AKR} were able to find a range
of values of the gravitational coupling that were compatible with the
observations of the primordial $D$, $^3${\it He}, $^4${\it He} and
$^7${\it Li} abundances. The range turned out to be rather large,
$\Delta G/G = 0.2$ at the $1\sigma$ level, due to the large statistical
and systematic errors of the observations.

This result will now be used to constrain the running of $G$ in an
asymptotically free theory of gravity. As mentioned above, in a theory
with a scale-dependent
gravitational constant, the maximum value of $G$ at a given time is the
one that corresponds to the physical horizon distance at that time.
During primordial nucleosynthesis, the horizon distance grows from a few
light-seconds to a few light-minutes, {\it i.e.} less than a few
milliparsecs. At that scale we find $\Delta G/G = 0.07$, see
Eq.~(\ref{DEL}), which is much less than the allowed variation of $G$
given in \cite{AKR}. Therefore, primordial nucleosynthesis does not rule
out the possibility of an asymptotically free gravitational coupling.
Of course, a light-second is about the distance to the Moon, and there
are similar constraints on a variation of $G$ at this scale coming from
lunar laser ranging, $\Delta G/G < 0.6$ \cite{TEGP}.
As a consequence, a theory where the gravitational constant varies more
quickly with scale would be ruled out by observations.

\subsection{Binary pulsars}

The precise timing of the orbital period of binary pulsars and, in
particular, of the pulsar PSR 1913+16, provides another way of obtaining
a model-independent bound on the variation of the gravitational coupling
\cite{TW,DGT}. Since the semimajor axis of that system is just
about a few light-seconds, the resulting limits on the variation of $G$
can be readly compared with the ones arising from nucleosynthesis.
The observational limits on the rate of change of the orbital period,
mainly due to gravitational radiation damping, together
with the knowledge of the relevant Keplerian and post-Keplerian orbiting
parameters, allows one to obtain the following limit \cite{TW,DGT}:
\begin{equation}
\sigma \equiv {\Delta G\over G} < 0.08\,h^{-1} \ ,
\end{equation}
where $h$ is the value of the Hubble parameter in units of 100 km/s/Mpc.
For $h = 0.5$, one obtains $\sigma = 0.16$ which is more stringent than
the nucleosynthesis bound, but is still compatible with the fit
(\ref{DEL}) for $r$ of a few light-seconds.

\subsection{Gravitational lensing}

Gravitational lensing of distant quasars by intervening galaxies may provide,
under certain assumptions, yet another method of constraining, on large
scales, the variability of the gravitational coupling. The four observable
parameters associated with lensing, namely, image splittings, time delays,
relative amplification and optical depth do depend on $G$, more precisely on
the product $GM$, where $M$ is the mass of the lensing object. This dependence
might suggest that limits on the variability of $G$ could not be obtained
before an independent determination of the mass of the lensing object.
However, as the actual bending angle is not observed directly, the relevant
quantities are the distance of the lensing galaxy and of the quasar. Since
these quantities are inferred from the redshift of those objects, they depend
on their hand on $G$, on the Hubble constant, $H_0$, and on the density
parameter, $\Omega_0$. However, as we have previuosly seen, a
scale-dependent gravitational coupling implies also a dependence on scale of
$H_0$ and $\Omega_0$, see Eqs.~(\ref{H2}) and (\ref{OME}).
This involved dependence
on scale makes it difficult to proceed as in Ref.~\cite{LENS},
where gravitational lensing in a flat, homogeneous and isotropic cosmological
model, in the context of a Brans-Dicke theory of gravity, was used to provide
a limit on the variation of $G$:
\begin{equation}\label{GLENS}
{\Delta G\over G} = 0.2 \ .
\end{equation}
Since for this limit $\Omega_0 = 1$ was assumed, while in a scale-dependent
model it is achieved via the running of the gravitational coupling, the
bound (\ref{GLENS}) contrains only residual variations of $G$ that have not
been already taken into account when considering the dependence on scale
of $H_0$ and $\Omega_0$. Of course, for models where the cosmological
parameters are independent of scale, the bound (\ref{GLENS}) can be readily
used to constrain the variability of $G$ on intermediate cosmological scales.
It is worth stressing that this method, besides being one of the few
available where this variability is directly constrained at intermediate
cosmological times between the present epoch and the nucleosynthesis era, it
is probably the only one which can realistically provide in the near future
even more stringent bounds on even larger scales by observing the lensing
of light from far away quasars caused by objects at redshifts of order
$z \geq 1$.

\subsection{Peculiar velocity field}

Since we expect the effects of a running $G$ to become important at very
large scales, one could try to explore distances of hundreds of Mpc,
where the gravitational coupling is significantly different from that of
our local scales. That is the realm of physical cosmology: peculiar
velocity fields and structure formation. Unfortunately, it is also the
realm of large observational uncertainties, which precludes any
reasonable detection of the effect we are looking for. However, with the
planed future sky surveys like the Sloan survey (SDSS) and powerful
telescopes like HST and Keck, one might expect this effect to become
observable in the not-so-far future. A possible signature would be a
mismatch between the velocity fields and the actual mass distribution,
such that at large scales the same mass would pull more strongly.
To be more specific, in an expanding universe there is a relation
between the kinetic and gravitational potential energy of density
perturbations known as the Layzer-Irvine equation (for a detailed
description see Ref.~\cite{Peebles}). It is a generalization of the
Newtonian Virial Theorem for non-liner gravitational instabilities
(although still non-relativistic), and can be written as a relation
between the mass-weighted mean square velocity $\bar v^2$ and the mass
autocorrelation function $\xi(r)$,
\begin{equation}\label{LI}
\bar v^2 (r) = 2\pi G~ \rho_b~ J_2(r)\ ,
\end{equation}
where $\rho_b$ is the mean local mass density and
$J_2(r) = \int_0^r r~dr~\xi(r)$. The galaxy-galaxy correlation function
can be parametrized by $\xi(r) \sim (r/r_0)^{-1.8}$ with $r_0 = 5 h^{-1}$
Mpc, while the cluster-cluster correlation function has the same
expression with $r_0 = 20 h^{-1}$ Mpc. This means that the velocity
field (\ref{LI}) should be proportional to $(r/r_0)^{0.2}$, unless the
gravitational constant has some scale dependence. So far the relation
seems to be satisfied, under huge experimental errors (for a review see
Ref.~\cite{Dekel}), except perhaps for a discrepancy in the motion of
the Local Group towards the Abell cluster, $150 h^{-1}$ Mpc away,
with velocities up to $v = 689 \pm 178$ km/s \cite{LP}, which might
indicate some anomaly in the velocity-mass relation. Unfortunately, the
errors are so large that it would be naive to infer from this a scale
dependence of $G$. Even worse, phenomenologically there is a
proportionality constant between the galaxy-galaxy correlation function and the
actual mass correlation function, the so- called biasing factor, which is
supposed to be scale dependent and could mimic a variable gravitational
constant. However, future sky surveys might be able to constrain more
strongly the relation (\ref{LI}) by measuring peculiar velocities with
better accuracy at larger distances. It might then be possible to extract
the scale-dependence of $G$.

\

Another very important area of cosmology in which bounds on a
hypothetical scale dependence of $G$ can be obtained in the forseable
future is the large scale structure of the Universe, i.e. theories of
structure formation and evolution. They deal with the largest possible
scales, all the way up to the horizon, and thus are presumably the
most likely to be sensitive to a strong scale dependence of the
gravitational constant. Unfortunately, as mentioned repeatedly before,
those are the regions with largest uncertainty errors. However, present
and future experiments like COBE, IRAS, the Sloan Digital Sky Survey
(SDSS), etc. will soon begin to constrain the existing models of
structure formation like cosmic strings, CDM, MDM, etc.\footnote{For a
recent review see Ref.~\cite{WSS}.} and therefore leave room for
a possible determination of the scale dependence of $G$. In particular,
as shown in Ref.~\cite{Bertolami}, the amplitude of linearised density
perturbations grow more quickly than in general relativity and could be
even more important in the non-linear regime, which might help
accelerate structure formation in an early epoch without the need of
introducing non-baryonic dark matter with {\it ad hoc} properties.
It is certainly worthwhile investigating the relevance of this effect
in explaining the large scale structure of the Universe.

\section{\label{5} Conclusions}

Let us summarise our results and comment on some further directions of
investigation. We have seen that the running of the gravitational
coupling is compatible with the observational fact that the rotation
curves of galaxies are constant provided some amount of baryonic dark matter
is allowed, actually about $44 \%$ less than what is required for a
constant $G$. This could also explain why we see less microlensing events
towards the halo than in the direction of the bulge of our galaxy.
Failure in reproducing the predicted distribution of baryonic dark
matter would signal either that the approach adopted here is unsuitable
or that the fit (\ref{DEL}) is inadequate, perhaps suggesting
alternative scale dependences like those discussed in the introduction.
For the purpose of
distinguishing between either of them, we have looked for possible
bounds on variations of $G$ with scale from primordial nucleosynthesis,
lunar laser ranging, variations in the period of binary pulsars,
macroscopic gravitational lensing and even deviations in the peculiar
velocity flows. Unfortunately, as observational errors tend to increase
with the scale probed, we cannot yet seriuosly constrain an increase of
$G$ with scale, as proposed by the asymptotically free theories of gravity.
Our study may provide nevertheless a guidance for further efforts in
constraining the variability of the gravitational coupling in other
cosmological or quantum gravity models.

\section*{ Acknowledgements}

O.B. would like to thank Vittorio de Alfaro for several interesting
discussions on the subject of this paper. J.G.B. thanks John Barrow,
Amitabha Lahiri, Andrew Liddle and David Wands for very useful comments
and for pointing out various relevant references. J.G.B. would also like
to thank the warm hospitality and financial support of the TH--Division
at CERN, where part of this work was performed.

\end{document}